\documentclass[%
12pt, 
 amsmath,amssymb,
 aps, physrev,
]{revtex4-2}

\usepackage{graphicx}
\usepackage{amsfonts}

\usepackage[colorlinks=true,linkcolor=blue,citecolor=blue,urlcolor=blue]{hyperref}
\usepackage[capitalise]{cleveref} 

\usepackage{bm}
\DeclareMathOperator{\tr}{tr} 
\DeclareMathOperator{\Str}{Tr} 
\DeclareMathOperator{\T}{T} 
\DeclareMathOperator{\D}{d} 
\DeclareMathOperator{\diag}{diag} 
\newcommand{\x}{{\bf x}}
\newcommand{\q}{{\bf q}}

\begin{document}

\title{Functional renormalization group equations for antisymmetric tensor field models at finite temperature}
\author{Georgii Kalagov}
\affiliation{Joint Institute for Nuclear Research, Joliot-Curie, 6, 141980 Dubna, Russia
}

\begin{abstract}
Within the framework of the functional renormalization group, we derived the flow equations for the scale-dependent effective action at finite temperature for models involving an antisymmetric rank-2 tensor field. The analysis focuses on scenarios where the vacuum expectation value emerges due to symmetry breaking patterns, specifically $SU(n) \to USp(n)$ and $SO(n) \to SU(n/2)$. The derived equations provide insights into the behavior of these models under varying scales and temperatures, contributing to the understanding of phase transitions in systems with complex symmetry structures.
\end{abstract}

\maketitle

\section{Introduction}
The study of effective field models--we consider bosonic ones--with enlarged symmetry has attracted considerable attention in condensed matter physics due to their versatility in describing a wide range of physical systems. These models are constructed following the Landau-Ginzburg-Wilson (LGW) formalism, wherein the geometric properties of collective fields, along with the symmetries of interactions, capture the essential characteristics of the systems under consideration. Depending on the model parameters, different symmetry-breaking scenarios can be realized, leading to distinct macroscopic phases.  Significant progress has been made in this field, including studies of the thermodynamic properties of $SU(n)$ fermionic gases \cite{Cazalilla2014,Gorshkov2014,Sonderhouse2020,Huang2023}. Recent studies on the critical properties of $SO(n)$ effective models are presented in \cite{Janssen22, Bonati24, Antonov2016, Lebedev2022}. The theoretical framework also extends to the models with $Sp(n)$ internal symmetry \cite{Ramires2017}.  A time-honored framework for the theoretical investigation of universal scaling in such systems is based on LGW $\phi^4$ models, which are perturbatively renormalizable in dimension $d = 4 - \varepsilon$, with critical exponents computed in the form of $\varepsilon$-expansions. The description of multicomponent systems extends beyond the investigation of a narrow critical region characterized by universal thermodynamic behavior. Calculating physical parameters, such as the heat capacity, the compressibility, and other relevant quantities \cite{ku} for Bose or Fermi liquids/gases, requires knowledge of the grand thermodynamic potential $\Omega(T, \mu)$ over a  broad range of its arguments and for arbitrary values of the coupling constants. The functional renormalization group (FRG),  see the review \cite{dupuis2021},  is well-suited for computing  such thermodynamic quantities, as it provides a non-perturbative and systematic way to account for fluctuations across all scales. This capability is crucial for predicting physical observables that depend on the specific details of  system interactions. Integrating the quantum and thermal fluctuations up to a given cutoff scale $k$ yields the initial value problem for the scale-dependent action $\Gamma_k[\phi]$
\begin{equation}
\label{eq:weq}
     \partial_k \Gamma_k[\phi] = \frac{1}{2}  \hat{\partial}_k \Str  \ln(\Gamma_k^{(2)}[\phi]+R_k), \quad  \Gamma_{k = \Lambda}[\phi] = S[\phi].
\end{equation}
The derivative $\hat{\partial}_k$ acts only on the cutoff function $R_k$, and $\Gamma_k^{(2)}[\phi]$ is the Hessian of $\Gamma_k[\phi]$. The classical action $S[\phi]$, which serves as the initial condition, parametrizes the model at the microscopic scale $\Lambda$. The function $R_k$ acts as a regulator that suppresses modes with momenta smaller than $k$. There is some flexibility in choosing the regulator, and one of the commonly used options is Litim's regulator \cite{dupuis2021}, given by $R_k(\omega, \q) = (k^2 - \q^2) \Theta(k^2 - \q^2)$.  By following the scale evolution towards $k = 0$, we derive the quantum effective action $\Gamma[\phi]=\Gamma_{k=0}[\phi]$, which contains all relevant universal and non-universal information. In particular, the grand thermodynamic potential is given by the minimum of $\Gamma[\phi]$, i.e., $\Omega = - T \min_{\phi} \Gamma[\phi]$. 

This work aims to present flow equations akin to \cref{eq:weq} for antisymmetric tensor field models at finite temperature, which have been discussed in the context of Cooper pairing in fermionic systems, as referenced in the aforementioned works. The classical LGW potential in such models is given by the expression $V_{\text{LGW}} = a + b\,   \tr(\phi^{\dag} \phi) +c_1 \tr(\phi^{\dag} \phi \phi^{\dag}  \phi) +  c_2 \left(\tr(\phi^{\dag} \phi)\right)^2+ \dots$. The field $\phi$ may be either complex or real; in the latter case, the conjugate transpose $\phi^{\dag}$ is replaced by its transpose $\phi^{\T}$. It can be interpreted as arising from the Hubbard-Stratonovich transformation of the four-fermion term, and fermion fluctuations generate a kinetic term for  $\phi$. The approximate solution of the equation \cref{eq:weq} is sought in the form of the gradient expansion, and to leading order, it is given by 
\begin{align}
 \label{eq:im}
 \Gamma_k & = \int\limits_{0}^{\beta} d {\tau} \int d^{\D} x  \, \Big(V_k(\phi^{\dag} \phi) + \tr({\bm \nabla} \phi^{\dag} {\bm \nabla} \phi) +   \tr(\phi^{\dag} \partial_{{\tau}} \phi)  \Big), &\text{for } \phi_{a b} \in \mathbb{C}, \\
 \label{eq:real}
  \Gamma_k & = \int\limits_{0}^{\beta} d {\tau} \int d^{\D} x\,   \Big(V_k(\phi^{\T} \phi) + \tr({\bm \nabla}\phi^{\T} {\bm \nabla} \phi) +   \tr( \partial_{{\tau}} \phi^{\T} \partial_{{\tau}} \phi)  \Big),  &\text{for  } \phi_{a b} \in \mathbb{R},  
\end{align}
where $\phi=\phi_{a b}({\tau},\x)$ represents the antisymmetric field of size $n \times n$ ($n$ is even), which is a function of the Euclidean time variable ${\tau} \in [0,\beta=T^{-1}]$ and spatial coordinates $\x\in \mathbb{R}^{\D}$, and satisfies the periodic boundary condition $\phi(0,\x)= \phi(\beta,\x)$.  The structure $\tr(\phi^{\T} \partial_{{\tau}} \phi)$ is absent in \cref{eq:real} due to the imposed boundary condition. Consequently, the leading nontrivial thermal contribution is given by the $\partial_{{\tau}}^2$ term. The local potentials $V_k$ depend on the field monomials that are invariant under the internal symmetry, and   $V_{k = \Lambda} = V_{\text{LGW}}$.
The equation for the potential is obtained by projecting \cref{eq:weq} onto a constant field: $\partial_k V_k = T \partial_k \Gamma_k[\phi]|_{\phi = \left\langle\phi\right\rangle} / \left((2\pi)^{\D} \delta^{\D}(0)\right)$. The tensor structure of the background field $\left\langle\phi\right\rangle$, defining the specific symmetry breaking scenario, is determined by the physical context. After choosing the scenario, the Hessian in \cref{eq:weq} must be computed to obtain the desired equations.

\section{\label{sec:sym}Aspects of symmetry}

\paragraph{Complex field case.}

The composite field $\phi$ in \cref{eq:im} transforms as $\phi \to g \phi  g^{\T}$, where $g \in U(n)$. A nonzero expectation value $\left\langle\phi\right\rangle$ leads to a nonvanishing anomalous Green function and induces a gap in the fermionic spectrum.  Its general form is $\left\langle\phi\right\rangle = \diag\{ \varphi_{1} \epsilon,\dots, \varphi_{k} \epsilon\}$, with the pairing amplitudes $\varphi_{1},\dots,\varphi_{k} \in \mathbb{C}$, $\epsilon$ is an antisymmetric unit tensor of size $2\times 2$, and $k=n/2$. If $k'$ of the total $k$ amplitudes are equal and the rest vanish, the unbroken symmetry is $SU(n-2k')\otimes USp(2k') $, corresponding to $k'$ gapped branches in the fermionic spectrum and $k-k'$ gapless states. In this work we consider the fully gapped superfluid state, i.e. $k'= k$, and all $\varphi$-amplitudes are equal. For computational convenience, we use the symplectic basis, where $\left\langle\phi\right\rangle = \varphi \, \Sigma / \sqrt{2 n}$ with $\varphi \in \mathbb{R}_+$ and
\begin{align}
\label{eq:vev}
 \Sigma  =  \begin{pmatrix}
 0 &I_{n/2} \\
 -I_{n/2} &  0  \\
\end{pmatrix},
\end{align}
thus the phase transition is accompanied by symmetry breaking $U(n) \to USp(n) = \{ h\in SU(n) \  | \ h  \Sigma h^{\T} =  \Sigma \}$. The dimension of the unbroken subgroup is $\dim USp(n) = n(n+1)/2$, see \cite{isaev}.

We denote the unbroken generators by $\Xi_{i}$, with $i = 1,\dots,  n(n+1)/2$. The set of remaining generators is $I_{n}$ and $\Pi_{a}$, where the unit matrix $I_{n}$ corresponds to $U(1)$ symmetry of global phase rotation, while $ \Pi_{a}$ are traceless hermitian, $a = 1,\dots,  n(n-1)/2-1$. The subgroup generators $\Xi_{i}$ satisfy $ (\Sigma \Xi_i)^{\T} = \Sigma \Xi_i$, while the broken generators, which span the Goldstone manifold $U(n)/USp(n)$, meet $(\Sigma \Pi_a)^{\T} = -\Sigma \Pi_a$, and are normalized as $\tr (\Pi_a \Pi_b) = \delta_{a b}/2$. Using these generators we can parameterize the field around the ground state as  
\begin{align}
 \begin{split}
     \phi &=  \exp\left(\frac{i \alpha}{\varphi \sqrt{2 n }}\right)  \exp\left(\frac{i \pi_a \Pi_a}{2 \varphi}\right) \Big( \left\langle\phi\right\rangle  +\frac{\sigma}{\sqrt{2 n }} \Sigma  + \xi_a \Pi_a \Sigma \Big) \exp\left(\frac{i \pi_a \Pi^{\T}_a}{2\varphi}\right)\\[1ex]  
     & \approx
      \frac{\varphi}{\sqrt{2 n }}   \Sigma + \frac{\sigma + i \alpha}{\sqrt{2 n }} \Sigma + (\xi_a  + i \pi_a) \Pi_a \Sigma.
\end{split}
\end{align}
Here $\sigma$, $\alpha$, $\bm{\xi}$ and  $\bm{\pi}$ are real-valued fields, and  the total number of real components is respectively $n_{\sigma} + n_{\alpha} +n_{\xi} + n_{\pi} \equiv 1 + 1 + [ n(n-1)/2 - 1] +  [n(n-1)/2 - 1] =  n(n-1)$, which equals the number of components of a complex antisymmetric matrix. The singlet $\alpha$ and $\bm{\pi}$-field are the ``Goldstone modes'', while $\sigma$ and $\bm{\xi}$ are the ``radial modes''. We chose the normalization factor in the denominators to obtain canonical kinetic terms for all modes, e.g., $(\partial_i \alpha)^2/2$. In the chosen basis, the Hessian is diagonal in the background field \cref{eq:vev}, making the right-hand side of \cref{eq:weq} easy to compute.

\paragraph{Real field case.}
For real-valued fields $g \in SO(n)$. In case the vacuum expectation value has the same form as above $\left\langle\phi\right\rangle = \varphi  \Sigma / \sqrt{2 n}$, the symmetry-breaking pattern is $SO(n) \to U(n/2)$. The generators of the unbroken subgroup $\Sigma$ and $ \Xi_{\alpha}$, $\alpha = 1,\dots, (n/2)^2-1$, satisfy $(\Xi_{\alpha} \Sigma)^{\T} = \Xi_{\alpha} \Sigma$, $\Xi_{\alpha}^{\T} = -\Xi_{\alpha}$, $\tr(\Xi_{\alpha} \Sigma)=0$, and have the block structure 
\begin{equation}
\label{eq:h}
\Xi  =  \begin{pmatrix}
 X & Y \\
 -Y &  X  \\
\end{pmatrix} \in {\mathfrak so}(n), \quad X^{\T} = - X, \quad Y^{\T} = Y, \quad  \tr Y = 0,  \quad Y + i X \in {\mathfrak su}(n/2).
\end{equation}
The remaining generators $\Pi_{\beta}$, $\beta = 1,\dots (n/2)^2 - n/2$ span the Goldstone manifold $SO(n)/U(n/2)$, and satisfy $ (\Pi \Sigma)^{\T} = - \Pi \Sigma$ and $\Pi^T = -\Pi$. They can be represented in the  block form
\begin{equation}
\Pi  =  \begin{pmatrix}
 P & Q \\
 Q &  -P  \\
\end{pmatrix}, \quad P^{\T}  = - P, \quad Q^{\T}  = - Q.
\end{equation}
Using these generators we can parameterize the field around the ground state as 
\begin{align}
 \begin{split}
    \phi &= \exp\left(\frac{\pi_{\beta} \Pi_{\beta}}{2 \varphi}\right) \Big( \left\langle \phi \right\rangle + \frac{\sigma}{\sqrt{2 n}} \Sigma  +  \xi_{\alpha} \Xi_{\alpha} \Big) \exp\left(-\frac{\pi_{\beta} \Pi_{\beta}}{2 \varphi}\right)   \\[1ex]  
    &\approx \frac{\varphi}{\sqrt{2 n}} \Sigma  +    \frac{\sigma}{\sqrt{2 n}} \Sigma+ \xi_{\alpha} \Xi_{\alpha} + \pi_{\beta} \Pi_{\beta} \Sigma.
    \end{split}
\end{align}
The normalization conditions $\tr (\Xi_{\alpha} \Xi_{\alpha'}) = -\delta_{\alpha \alpha'}/2$ and $\tr (\Pi_{\beta} \Pi_{\beta'}) = -\delta_{\beta \beta'}/2$ are imposed.  The total number of real components ${\bm \xi}, {\bm \pi}$, and $\sigma $  is $n_{\xi} + n_{\pi} + n_{\sigma } \equiv  [(n/2)^2-1] + [(n/2)^2-n/2] + 1 =  n(n-1)/2$.  

\section{Results}
Given the property of the background field $\Sigma^{\T} \Sigma = I_n$, it is convenient to choose invariants for expressing the potential $V_k$ in the following manner. The quadratic invariant is defined as $\rho = \tr(\phi^{\T}\phi)$. Higher-order invariants are constructed from the powers of the traceless tensor $\phi^{\T}\phi - I_n \rho/n$ (for the complex field case $\T \to \dag$). Specifically, the quartic invariant is represented by $\vartheta = \tr\left(\phi^{\T}\phi - I_n \rho/n\right)^2$. Such invariants evaluated on the background field are identically zero. In a vicinity of \cref{eq:vev}, the potential can be presented as a series in higher-order invariants $V_k(\phi^{\T}\phi) = U_k(\rho) + W_k(\rho) \vartheta   + \text{higher-order terms}$. The flow equation for the potential  $U_k$ we obtain is  
\begin{equation}
    \partial_k U_k =    \frac{T}{2}\sum_{\omega \in 2 \pi T \mathbb{Z}}\int\frac{d^{\D} q}{(2\pi)^{\D}} \sum_{a = \pi, \sigma, \xi}  \frac{  n_{a} \,\partial_k R_k(\omega, \q)}{  (\omega^2 + \q^2 + R_k(\omega, \q) + m_{a}^2)}.
\end{equation}
The eigenvalues of the ``mass'' matrix are $m_{\pi}^2 =  U_k'$,  $m_{\sigma}^2 =  U_k' + 2 \rho  U_k''$, and  $ m_{\xi}^2 =  U_k' + 4 \rho  W_k/n$, where $ U_k' \equiv \partial_{\rho}U_k$. For Litim's cutoff function $R_k$ the momentum integral and the discrete frequency sum can be computed analytically, and the flow equations can be expressed in the form
\begin{equation}
\label{eq:u}
    \partial_k U_k =   \sum_{a = \pi, \sigma, \xi}  n_{a} \, h_1(E_{a}), \quad h_1(x)  =  \frac{k^{d+1}}{2  {x} } \coth\left(\frac{ {x}}{2 T} \right) \frac{(4\pi)^{-\D/2}}{\Gamma(\D/2+1)},
\end{equation}
where $E_{a} = \sqrt{k^2+ m_{a}^2}$. This equation depends on $W_k$ through $m_{\xi}^2$ and is therefore not closed. To obtain the flow for $W_k$, the structure of the background field \cref{eq:vev} is insufficient; the $\left\langle\phi\right\rangle$  must be infinitesimally extended in some direction $\Sigma_{\perp}$, such that $\tr (\Sigma_{\perp} \Sigma) = 0$. Subsequently, algebraic constructions similar to those in \cref{sec:sym} must be performed to compute the Hessian. Omitting subtle details and cumbersome calculations,  we present the flow equation
\begin{align} 
\begin{split}
\label{eq:w}
  \partial_t  W_k =&  \frac{(n-4) }{2} W_k^2 h_3(E_{\pi})+ 
 \frac{9 (n^2-16)}{2 n }  W_k^2 h_3(E_{\xi}) + \frac{n^2}{
 8 \rho} W_k  \big( h_2(E_{\xi}) - h_2(E_{\pi})   \big)   \\[1ex] 
 &-  \frac{n^2+12}{4}\, W_k' \, h_2(E_{\xi}) - \frac{n (n-2)}{
 4}  W_k' \, h_2(E_{\pi})  -  (2\,{ W_k''}\,\rho+5\,{ W_k'})  h_2(E_{\sigma}) \\[1ex]  
  & + \left( \frac {n U'' - 2 W_k}{2 \rho}-\,\frac { \left( n U''+4\,{W_k'}\,\rho+4\,W_k \right) ^{2}}{ n \left(m_{\sigma}^2
-m_{\xi}^2 \right)}  \right) \big(  h_2(E_{\sigma}) - h_2(E_{\xi}) \big), 
 \end{split}
\end{align}
where the higher-order threshold functions can be computed recursively $ h_{l+1}(x)  = - \partial_x h_l(x)/ (2 l x)  $. Despite the apparent singularity of the equation at $\rho = 0$ or $E_{\sigma} = E_{\xi}$, the right-hand side is well defined at these points, since $\lim_{\rho \to 0} [h_2(E_{\xi}) - h_2(E_{\pi})]/\rho \sim W h_3(E_{\pi})$. Similarly, $\lim_{m_{\sigma}^2 \to m_{\xi}^2} [h_2(E_{\sigma}) - h_2(E_{\xi})]/[m_{\sigma}^2 - m_{\xi}^2] \sim h_3(E_{\xi})$. 

The system of obtained equations should be solved with the initial conditions $U_{\Lambda}$ and $W_{\Lambda}$, which can be directly obtained from the LGW classical action or fermionic loop. Following this, one can reveal possible phase transitions and compute the thermodynamic parameters. For instance, the pressure $p = -\Omega/\mathcal{V} = \min_{\rho} U_{k=0}(\rho)$, the specific entropy $s = \partial p /  \partial T$, and so on.

\paragraph{Classical regime $T \to \infty$:} The analysis of critical behavior reduces to determining the leading infrared asymptotics, which are characterized by massless propagators, indicating that the primary contribution arises from the zero Matsubara frequency. This regime can be formally obtained by taking the limit $T \to \infty$ in the thermal flow equations. In this limit, the threshold functions simplify to $h_l(x) \sim  T x^{-l}$. Notably, to ensure a proper transition to this limit, the rescalings $U_k \to T U_k$, $\rho \to T \rho$, and $W_k \to W_k/T$ must be applied. In the case of  $n = 2 $, a real antisymmetric matrix is parameterized by a single real number, and \cref{eq:u} reduces to the well-known equation for the $\mathbb{Z}_2$ symmetric model. For $n = 3$, although we consider even values of $n$, the model reduces to the $\mathrm{SO}(3)$ symmetric theory, for which the functional FRG equations are also known \cite{dupuis2021}. For $n = 4$, due to the isomorphism $\mathfrak{so}(4) \cong \mathfrak{so}(3) \oplus \mathfrak{so}(3)$, the model reduces to the $MN$-model \cite{mn}, where the order parameters are two three-component vectors. This model possesses a nontrivial infrared attractive fixed point, and the system exhibits critical scaling. Within the perturbative expansion of the system \cref{eq:u,eq:w} in dimension $ d = 4 - \varepsilon$, the leading term coincides with that of the renormalization group functions obtained by the authors of \cite{Antonov2016, Lebedev2022}. Dimensional analysis of the equations shows that the solution can be represented in the form $U_k(\rho) = T k^{d} F_u(\rho / (T k^{d-2}), t)$ and $W_k(\rho) = T^{-1} k^{4-d} F_w(\rho / (T k^{d-2}), t)$, where $F_u, F_w$ are dimensionless functions, and $t = -\ln(k / \Lambda)$. From this, the canonical dimensions of all quantities also follow.

\paragraph{Quantum limit $T \to 0$:} In the flow equations,  we can safely take the zero-temperature limit. The same result would be obtained if the model were initially formulated at zero temperature. However, in this case, the integration over the ``time'' interval  $[0, \beta]$ must be extended to $[-\beta, \beta]$, taking into account the periodic properties of the bosonic field. Only after this extension can one pass to the limit $\beta \to \infty$, because the theory on the time semi-axis differs significantly from the theory on the entire axis.

\begin{figure}[t!]
    \includegraphics[height=6cm]{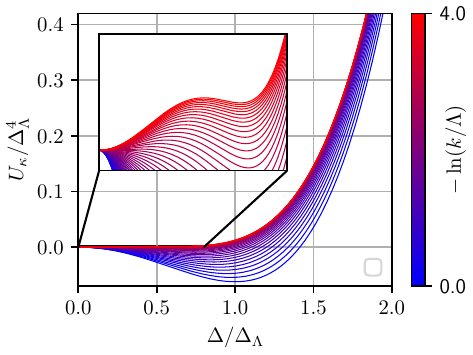}
    \includegraphics[height=5.9cm]{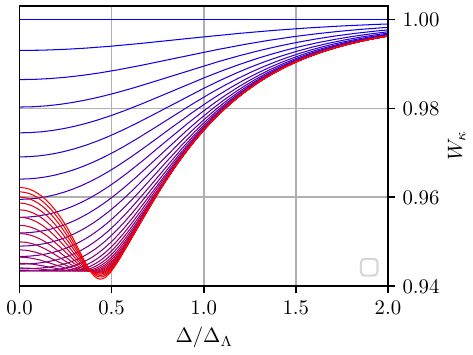}
    \caption{Evolution of the functions $U_k$ and $W_k$ with respect to the scale $k$. Starting from the constant value $W_{\Lambda}$, the function $W_k$ develops a nontrivial shape. The graphs are depicted for $n=4$, $\lambda^{(1)}_{\Lambda} = \lambda^{(2)}_{\Lambda} = 1$, and $\Delta_{\Lambda} = 0.56 \Lambda$.}
    \label{fig:flow}
\end{figure}

Consider a typical solution of system \cref{eq:u,eq:w} in this regime. As initial conditions, we take the model example $U_\Lambda(\rho) = \lambda^{(1)}_{\Lambda} (\rho - \rho_{\Lambda})^2 /2$ and $W_\Lambda(\rho) = \lambda^{(2)}_{\Lambda}$, where $\rho \equiv \Delta^2/2$, and $\Delta_{\Lambda} = \sqrt{2 \rho_{\Lambda}}$ represents the condensate value at zero temperature.
A global numerical solution reveals the fluctuation-induced first-order quantum phase transition \cref{fig:flow} as the magnitude of $\Delta_{\Lambda}$ changes.  The fluctuations lead to a nontrivial shape of the coefficient function $W_k$ in the invariant expansion. Although the full effective action is a convex functional, the obtained solution is not. This is primarily due to the use of an ansatz, which truncates the space for constructing the derivative expansion. A secondary reason is the inability to reach the $k = 0$ limit in the numerical scheme. The smallest value of $k_{\min}$ at which we stop the flow, in the \cref{fig:flow} $k_{\min}\approx \Lambda \exp(-4)$, is chosen to  ensure stability of the solution under further reduction of $k$. 
The invariant expansion of $V_k(\phi^{\T}\phi)$, similar to the Dyson-Schwinger hierarchy, generates higher-order invariant contributions in the equation for $W_k$. Preliminary estimates suggest that these contributions are negligible for determining phase transition parameters. Notably, the simple Taylor series approach to solve equations like \cref{eq:u,eq:w}, effective for continuous phase transitions, is unsuitable for discontinuous ones. This is because the position of the minimum, around which the expansion is made, is {\it a priori} unknown, and the magnitude of the order parameter jump is generally not small. The FRG system we derived is a Cauchy problem, with the initial condition for the potential defined over the region $\Delta \in (-\infty, \infty)$. Thus, there are no real boundary conditions at spatial infinity in a non-compact domain, and the problem is fully determined by the asymptotics of the flow equation and the initial condition. However, this cannot be directly implemented on a finite computational domain, necessitating one of two methods. The first method involves mapping the axis $(-\infty, \infty)$ to a compact domain, such as $[-1, 1]$, through an appropriate transformation. The second approach, used here, is to restrict the original domain by introducing a cutoff $\Delta_{\max}$. In this case, the emergent boundary conditions at the endpoints of the interval $[-\Delta_{\max}, \Delta_{\max}]$ must be consistent with the initial condition to make the problem well-posed. The numerical value of $\Delta_{\max}$ should be chosen to be significantly larger than the location of the potential minimum. For each specific case, it is necessary to verify which value of $\Delta_{\max}$ is sufficient to ensure the stability of the computed results. For example, in the case shown in \cref{fig:flow}, we use $\Delta_{\max}/ \Delta_{\Lambda} \sim 10$.  Note also, in the quantum limit, dimensional analysis shows that $U_k(\rho) = k^{d+1} f_u(\rho / k^{d-1}, t)$ and $W_k(\rho) =  k^{3-d}  f_w(\rho /  k^{d-1}, t)$, where $ f_{u},  f_{w}$ are dimensionless functions.

In these notes, we presented the functional renormalization group equations for antisymmetric tensor field models at finite temperature and illustrated their solution through a model example. For complex fields, the resulting thermal equations are too lengthy to include here. A detailed analysis of these equations in the context of the superfluid phase transition in large-spin fermionic systems is published in \cite{kalagov2025}. 

\label{sec:acknowledgement}
\section*{Acknowledgement}
The author thanks A.P. Isaev for the book \cite{isaev}, which provided valuable insights and guidance for this work. 

\label{sec:funding}
\section*{Funding}
The author  was supported by the Foundation for the Advancement of Theoretical Physics and Mathematics ``BASIS'' 22-1-4-34-1.

\bibliographystyle{apsrev4-1}
\bibliography{biblio}

\end{document}